\documentclass[12pt,preprint]{aastex}

\slugcomment{---}

\shorttitle{Hidden kinetic power of radio jets}
\shortauthors{Ito et al.}

\begin{document}



\title{The estimate of kinetic powers of jets in FRII radio galaxies:
existence of invisible components?}


\author{Hirotaka Ito\altaffilmark {1}, Motoki Kino\altaffilmark{2},
    Nozomu Kawakatu\altaffilmark{3},  Naoki  Isobe\altaffilmark{4},
    and
    Shoichi Yamada\altaffilmark{1,5} }





\altaffiltext{1}{Science and Engineering, Waseda University, 3-4-1 Okubo,
Shinjuku, Tokyo 169-8555, Japan}
\email{hito@heap.phys.waseda.ac.jp}
\altaffiltext{2}{ISAS/JAXA, 3-1-1 Yoshinodai, 229-8510 Sagamihara, Japan}
\altaffiltext{3}{National Astronomical Observatory of Japan, 181-8588 Mitaka, Japan}
\altaffiltext{4}{Cosmic Radiation Laboratory, 
  Institute of Physical and Chemical Research,
    Wako, Saitama, Japan 351-0198}
\altaffiltext{5}{Advanced Research Institute for Science \&
Engineering, Waseda University, 3-4-1 Okubo,
Shinjuku, Tokyo 169-8555, Japan}


\begin{abstract}

We 
investigate the total kinetic powers ($L_{\rm j}$) and ages 
($t_{\rm age}$) of powerful jets in FR II radio galaxies
by  comparison of the dynamical model
of expanding cocoons with observations.
We select four FR II radio sources (Cygnus A, 3C 223, 3C 284, 
and 3C 219), for which the mass-density profiles 
of intracluster medium (ICM) are known in the literature. 
It is found that large fractions $\gtrsim 0.02 - 0.7$
 of the Eddington luminosity
($L_{\rm Edd}$)
 are carried away as a  kinetic power of jet.
The upper limit of estimated $2 L_{\rm j} / L_{\rm Edd}$ 
are larger than unity
($\lesssim 10$)
for some sources,  suggesting a possibility of
super-Eddington mass accretions.
As a consequence of the large powers,
we also find that the total
energy stored in the cocoon ($E_{\rm c}$)  exceeds the
energy derived 
from the minimum energy condition for 
the energy
 of radiating non-thermal 
electrons and magnetic fields ($E_{\rm min}$): 
$4< E_{\rm c}/E_{\rm min} <310$. 
This implies that
most of the energy in cocoon is carried by invisible components such as thermal
 leptons (electron and positron) and/or protons.

\end{abstract}





\section{INTRODUCTION}

Relativistic jets in active galactic nuclei (AGN)
are a fundamental aspect of plasma accretion onto 
supermassive black holes (SMBHs).
Although the formation mechanism of relativistic jets
remains a longstanding problem, it is well established that 
they carry away some fractions of the available accretion power 
in the form of collimated beam
(e.g., Begelman et al. 1984 for review).
The total kinetic powers of AGN jets $L_{\rm j}$
is one of the most basic physical quantities characterizing the
jet.
A lot of authors have investigated
$L_{\rm j}$ in various ways so far
\citep[e.g.,][]{RS91,  CF93, WRB99}. 
It is, however, difficult to estimate $L_{\rm j}$,
since most of the observed emissions from AGN jets
are of non-thermal electron origin
and it is hard to detect the electromagnetic signals
from the thermal and/or proton components.
Hence, the free parameter describing the amount of 
the invisible plasma components always lurks in  
the estimates of $L_{\rm j}$ 
based on the non-thermal emissions.

The estimate of $L_{\rm j}$ 
for low-power Fanaroff-Riley I (FRI) radio galaxies 
has been motivated by the observations of ``X-ray cavity" %
which is the region embedded in ICM
with the suppressed X-ray surface brightness
and coincides with the  radio lobe
 \citep{BVF93}.
The cavities (or cocoons) are supposed to be a direct evidence of the
displacement of the ambient ICM by the shocked jet matter. 
Dynamical models of cavities are a good tool,
since the invisible plasma components as well 
as non-thermal electrons play a role for 
expansions of cavities.
For FR I sources, the total kinetic energy of the jet 
has been estimated as
\begin{eqnarray}
L_{\rm j}t_{\rm age} \sim 
\frac{\hat{\gamma}_{\rm c}}{\hat{\gamma}_{\rm c}-1}
P_{\rm c} V_{\rm c},  \nonumber
\end{eqnarray}
where 
$t_{\rm age}$, 
$P_{\rm c}$, 
$V_{\rm c}$,
$\hat{\gamma}_{\rm c}$,
are 
the source age,
the pressure,
the volume, 
the adiabatic index of the cavity,
respectively
\citep{FCB02, ADF06}.
In these studies, however,
the thermal pressure of surrounding ICM ($P_{\rm ICM}$)
is substituted 
for the one in the cavity (i.e., $P_{\rm c}\sim P_{\rm ICM}$).
This assumption may be applied only to subsonic expansions.

On the other hand, the cocoon pressure of powerful  
Fanaroff-Riley class II (FRII) radio galaxies is 
expected to be larger than that of 
the surrounding ICM,
 which is expressed as $P_{\rm c} > P_{\rm ICM}$
 (Begelman \& Cioffi 1989 hereafter BC89), and 
the cocoon of FR II radio galaxies
is likely to be expanding super-sonically. Then the substitution of 
ICM pressure for the cocoon pressure is not justified.
A new estimate 
of $L_{\rm j}$ for FR II radio galaxies by use of the 
dynamical model of cocoon expansions 
is proposed by \citet{KK05} (hereafter KK05), in which
$L_{\rm j}$ and  $t_{\rm age}$ are derived
from the comparison of the cocoon model with the 
morphology of the cocoon obtained by radio observations.
It should be stressed that  $P_{\rm c}$ is not assumed but solved
in this model. Hence it can be 
applied even to the cocoons with $P_{\rm c}>P_{\rm ICM}$.
So far, however, this estimate of $L_{\rm j}$  has been
applied only to Cygnus A.
The expansion of the number of samples 
is evidently crucially important for exploring general characteristics of the 
powerful AGN jets. 
For this purpose, we apply the method of KK05 to other 
bright FR II radio galaxies, for which the physical conditions of the
associated ICM  have been estimated in the literature. 

In the present work, we especially focus on the ratio of
$L_{\rm j}/L_{\rm Edd}$, where $L_{\rm Edd}$ is the 
Eddington luminosity of AGN, since $L_{\rm j}/L_{\rm Edd}$
is a more fundamental quantity than $L_{\rm j}$ from the point of view of the jet formation physics.
Another interesting quantity we examine in this work
is the ratio of the internal energy deposited in the cocoon ($E_{\rm c}$)
to the  minimum energy ($E_{\rm min}$)
obtained by the minimum energy condition
for radiating non-thermal electrons and magnetic fields (e.g., Miley 1980).
Some of the previous works studying
this quantity
(e.g., Hardcastle \& Worrall 2000; Leahy \& Gizani 2001)
reported that the cocoon pressure expected from the inferred $E_{\rm min}$
is smaller than the pressure of ambient matter, suggesting the
difference between $E_{\rm c}$ and $E_{\rm min}$.
due to the lack of minimum pressure 
against the pressure of ambient medium. 
Although these studies obtained the lower limit of the ratio,
it is the value of $E_{\rm c}$ that is of greater importance.

According to a large sample of
galaxies collected recently, the fraction of AGNs in all
the galaxies
is suggested to be $\sim20 \-- 40\%$, larger than previously thought
\citep{KHT03, MNG03}.
The interest in 
AGNs is gaining momentum in the 
context of the co-evolution of galaxies and 
their central black holes 
(e.g., Kawakatu et al. 2003, Granato et al. 2004;
Di Matteo et al. 2005).
AGN outflows in particular
are likely to be a key ingredient
in this context 
(Silk \& Rees 1998; Fabian 1999; King 2003).
The AGN feedback by the 
outflows may be also promising to explain 
the tight correlations between 
the  ratio of the mass of SMBH ($M_{\rm BH}$) to that of
 galactic bulge
(Kormendy \& Richstone 1995; Magorrian et al. 1998)
and the ratio of $M_{\rm BH}$  and the
stellar velocity-dispersion in the bulge
\citep{FM00, TGB02}.
In this sense, a robust estimate of the basic 
quantities such as $L_{\rm j}/L_{\rm Edd}$ and $t_{\rm age}$ of
radio loud AGNs at low $z$ is an important first step
for understanding the AGN feedback processes in the universe.

The outline of the paper is as follows. 
In \S2, the model of the expanding cocoon 
by KK05 is briefly reviewed. 
 In \S3, we explain how to extract the key quantities from the observations of four nearby 
FR II radio galaxies, Cygnus A, 3C 223, 3C 284, and 3C 219, which are
required for the comparison with the model.
We then estimate  the total kinetic power, $L_{\rm j}$,
and the dynamical ages, $t_{\rm age}$,
in  \S4.
 Finally in \S5, we summarize our results and
 discuss some implications on
 the physics of AGN jet. 
Throughout the paper, we adopt a cosmology with $H_0=71~{\rm km~s^{-1}}$,
 $\Omega_{\rm M}=0.3$, and  $\Omega_{\rm \Lambda}=0.7$
\citep{SVP03}. 

\section{COCOON MODEL}

\subsection{Basic equations}
\label{beq}

\begin{figure}[ht]
\begin{center} 
\includegraphics[width=8cm]{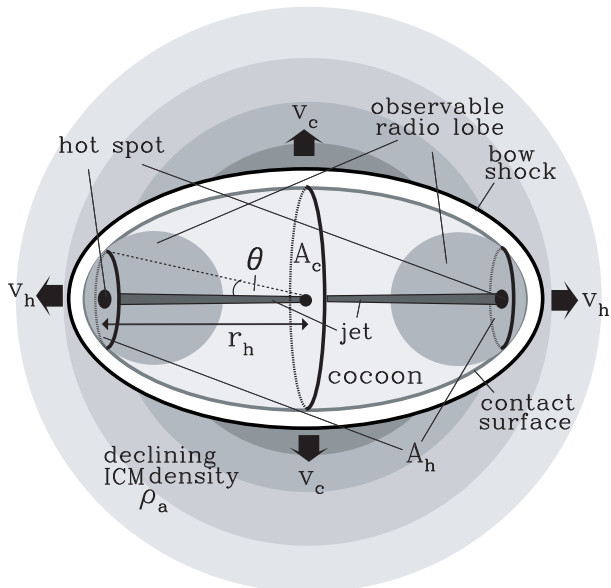}
\caption
{A cartoon of the employed model.  
The relativistic jet from FR II radio galaxy
interacts with
ICM with a declining density. 
Most of the kinetic energy
of jet is deposited in the cocoon, which is then
inflated by the internal energy.}
\label{cocoon}
\end{center}
\end{figure}

Based on BC89 and KK05,
we briefly summarize the cocoon model we employ in the following.
 We focus on the cocoon expansion in the over-pressured regime,
 namely $P_{\rm c} > P_{\rm a}$, where $P_{\rm c}$
and $P_{\rm a}$ are the  pressures of cocoon 
and  ambient ICM, respectively. 
We approximately describe the expansion of cocoon by the following
three equations:
(\ref{jet-axis}) the equation of the motion along the jet axis,
(\ref{lateral}) the equation for the sideways expansion, and 
(\ref{energy}) the energy equation.
They are expressed, respectively, as
\begin{eqnarray}
\frac{L_{\rm j}}{v_{\rm j}}=
\rho_{\rm a}(r_{\rm h})v_{\rm h}^{2}(t)A_{\rm h}(t),
\label{jet-axis}
\end{eqnarray}
\begin{eqnarray}
P_{\rm c}(t)=
\rho_{\rm a}(r_{\rm c}) \
v_{\rm c}(t)^{2}  ,
\label{lateral}
\end{eqnarray}
\begin{eqnarray}
\frac{dE_{\rm c}(t)}{dt}
  + P_{\rm c}(t)\frac{dV_{\rm c}(t)}{dt}
= 2 L_{\rm j}    ,
\label{energy}
\end{eqnarray}
where 
$v_{\rm j}$,
$\rho_{\rm a}$, 
$v_{\rm h}$,
$v_{\rm c}$, and
$A_{\rm h}$ 
  are 
the velocity of  jet,
the density of ambient medium,
 the advance velocity of cocoon head,
 the velocity of sideways expansion, and
 the cross sectional area of cocoon head, respectively.
Here 
 $E_{\rm c}=P_{\rm c} V_{\rm c}/({\hat{\gamma_{\rm c}}}-1)$
 is the total internal energy deposited in the cocoon, where
  $\hat{\gamma}_{\rm c}$  is 
 the specific heat ratio of the plasma inside cocoon. 
 The cocoon shape is approximated
 as a spheroid, and its volume is given as
 $V_{\rm c}(t)= (4 \pi /3)  r_{\rm c}(t)^2 r_{\rm h}(t)$. 
The distance from the jet apex to the hot spot and
 the radius of  cocoon body
 are obtained from
 $r_{\rm h}(t)=\int_{t_{\rm min}}^{t} v_{\rm h}(t')dt'$ and 
 $r_{\rm c}(t)=\int_{t_{\rm min}}^{t} v_{\rm c}(t')dt'$, respectively,
 and  $t_{\rm min}$ is the initial time of source evolution.
 Throughout this paper, we assume  $\hat{\gamma}_{\rm c} = 4/3$,
 since the cocoon is expected to be dominated by relativistic particles
 \citep{KKI07}.
 In these equations we also assume that 
 the jet has a relativistic velocity ($v_{\rm j} \sim c$) and
 that $L_{\rm j}$ is constant in time.
 The mass density 
 of ICM, $\rho_{\rm a}$,  is assumed to be given by
 $\rho_{\rm a}(r)
 ={\bar\rho}_{\rm a}(r/r_{0})^{-\alpha}$,
 where $r_{0}$ and 
 $\bar{\rho}_{\rm a}$ are the
 reference position and
 the ICM mass density at $r_{0}$, respectively.
 We set $r_{0}$ to be $r_{\rm h}(t_{\rm age})$,
 where  $t_{\rm age}$ is the present age  of cocoon,
 throughout this paper. 
 A cartoon of the cocoon model is illustrated in Fig. \ref{cocoon}.
 In this paper, we have slightly improved the model of BC89 and KK05
 as follows:
 (i) a more accurate
 definition of $V_{\rm c}$
  is employed,
 and
 (ii) the $PdV$ work, which is done by the cocoon against
 the contact discontinuity between the cocoon and the shocked ambient
  medium, is taken into account. 
 These corrections are necessary in the following quantitative estimate
 of $L_{\rm j}$. In fact, the estimated power is reduced by a factor of
 $\sim 50$ 
 from the value in KK05 for Cygnus A after taking account
 of the corrections (see \S \ref{improve} for details). 

 The model parameters are 
 $L_{\rm j}$ and $t$,
 and the
 unknown physical quantities are
 $v_{\rm h}$, 
 $v_{\rm c}$,
 $P_{\rm c}$, and
 $A_{\rm h}$. 
 Since the number of unknown  quantities is four,
 while that of
 basic equations is three,
 an additional condition
 is required
 for the system of equations to be closed.
 Here we assume that the cross sectional area of cocoon body
 $A_{\rm c} = \pi r_{\rm c}^2$ is given by 
 $A_{\rm c}(t) \propto t^{X}$ and treat
 $X$ as a free parameter which is determined by the imposed condition.
 Once the value of $X$ is determined,
 we obtain  $v_{\rm h}$, 
 $v_{\rm c}$,
 $P_{\rm c}$, and
 $A_{\rm h}$ 
  as a function of $L_{\rm j}$ and $t$.
 It is worth noting that 
 the model  is capable of producing various dynamics by tuning
 the value of $X$.
 For example, 
 the results of 2D relativistic 
 hydrodynamical simulations by Scheck et al. (2002)
 were reproduced fairly well in Kawakatu \& Kino (2006).


\subsection{Analytical solution}
\label{Ana}

%
%
 We  assume that the physical quantities  have 
 a power-law time-dependence of the form
 $A={\bar A} ~ (t/t_{\rm age})^{Y}$, where
 $Y$ is the power-law index.
 Then the index and
 coefficient 
 of $v_{\rm c}$, for example,  are determined as 
\begin{eqnarray}\label{v_c}
v_{\rm c}(t)=
{\bar v}_{\rm c}
\left(\frac{t}{t_{\rm age}}\right)^{0.5X-1}
=
\frac{{\bar A}_{\rm c}^{1/2}}{t_{\rm age}}
{\cal C}_{\rm vc}
\left(\frac{t}{t_{\rm age}}\right)^{0.5X-1} .
\end{eqnarray}
 From this relation and Eqs. (\ref{jet-axis})-(\ref{energy}),
 we obtain the following expressions:
\begin{eqnarray}
P_{\rm c}(t)=
{\bar\rho}_{\rm a} {\bar v}_{\rm c}^{2}
{\cal C}_{\rm pc}
\left(\frac{{\bar v}_{\rm c}}{v_{0}}\right)^{-\alpha}
\left(\frac{t}{t_{\rm age}}\right)^{X(1-0.5\alpha)-2},
\end{eqnarray}
\begin{eqnarray}\label{v_h}
v_{\rm h}(L_{\rm j},t)= 
\frac
{L_{j}}
{{\bar\rho}_{\rm a}{\bar v}_{c}^{2}{\bar A}_{\rm c}}
{\cal C}_{\rm vh} 
\left(\frac{{\bar v}_{\rm c}}{v_{0}}\right)^{\alpha}
\left(\frac{t}{t_{\rm age}}\right)^{X(-2+0.5\alpha)+2},
\end{eqnarray}
\begin{eqnarray}\label{ah}
A_{\rm h}(L_{\rm j},t)=
\frac{L_{\rm j}}
{v_{\rm j}{\bar\rho}_{\rm a} 
{\bar v}_{\rm h}^{2}}
{\cal C}_{\rm ah}
\left(\frac{{\bar v}_{\rm h}}{v_{0}}\right)^{\alpha}
\left(\frac{t}{t_{\rm age}}\right)
^{X(\alpha-2)(-2+0.5\alpha)+3\alpha-4} ,
\end{eqnarray}
 where 
 ${\cal C}_{\rm vh}= 
 0.75  (\hat{\gamma}_{c}-1)(0.5X)^{-\alpha}
 [3-(2-0.5\alpha)X] / [X(-1+0.5\alpha)(\hat{\gamma}_{c}-1) +
                         3\hat{\gamma}_{c} - 2]$, 
 ${\cal C}_{\rm vc}=0.5X/ \pi^{1/2}$,
 ${\cal C}_{\rm pc}=(0.5X)^{\alpha}$, and 
 ${\cal C}_{\rm ah}=[X(-2+0.5\alpha)+3]^{-\alpha}$, and
 $v_{0}\equiv r_{\rm h}(t_{\rm age})/t_{\rm age}$
 corresponds to
 the head speed assumed to be constant in time. 
 The difference in 
 ${\cal C}_{\rm vh}$ obtained here from that in KK05
 is due to 
 the correction made in Eq. (\ref{energy}).
 We assume 
 the conditions of $0.5X>0$ and $X(-2+0.5\alpha)+3>0$, 
 which ensure that the contribution at  $t_{\rm min}$ 
 to the  integrations of  $r_{\rm h}$ and  $r_{\rm c}$
 is small enough.
 The cases that we focus on in \S 3 
 clearly satisfy these conditions.

\subsection{Determination of $X$}

 As mentioned in \S\ref{beq},
 an additional condition that  
 determines the free parameter $X$ is required
 for the system of equations to be closed.
 In the pioneering study of BC89,  $A_{\rm h}(t) = {\rm const}$
 was assumed.
 However, as can be  confirmed from  numerical simulations
 \citep[e.g.,][]{CB92,  SA02},
 it is obvious that this condition is
 unlikely to hold for long-term evolutions
 from pc to Mpc scales.
 In this paper, we consider the following two conditions, which seem more
 reasonable and equally possible,
 in determining $X$:

\begin{itemize}
\item  {\bf Constant aspect ratio} (Case I)
\newline 
 The aspect ratio of
 cocoon, ${\cal R} = r_{\rm c}/r_{\rm h}$, is
 assumed to be constant  in time.
 This corresponds to the 
 widely-discussed 
  self-similar evolution \citep{B96, KF97, KA97, BDO97}.
  Since the time dependence of ${\cal R}$ 
  is given by
  ${\cal R}(t) \propto t^{[X(2.5-0.5\alpha)-3]}$, we obtain 
  $X=6/(5-\alpha)$ in this case.

%
%

\item  {\bf Constant opening angle} (Case II)
\newline
 The  opening angle,
 $\theta = {\rm tan}^{-1}( A_{\rm h}^{1/2}/ \pi^{1/2}r_{\rm h})$
 (see Fig. \ref{cocoon}), 
 is assumed to be constant in time.
 Although there has been no previous work that has employed
 this condition, it seems to be reasonable
 when the jet is precessing with a constant pitch angle.
 Since the time dependence of  ${\rm tan}\theta$
  is given by
  ${\rm tan}\theta(t) \propto t^{[0.25X(\alpha -4)^{2}+1.5\alpha-5]}$,
  $X= (20-6\alpha)/(4-\alpha)^2$ is obtained for this case.

\end{itemize}
%

  It should be emphasized here that these two independent conditions
 lead to the solutions that describe quite similar dynamical
 evolutions as long as the range of $\alpha$ listed in Table \ref{tab1} 
 is adopted ($1\leq \alpha \leq 2$).
 This can be seen as follows. If the constant opening
  angle is imposed (Case II), the evolution of the aspect 
 ratio is obtained as ${\cal R}(t)\propto t^{(2-\alpha)/(4-\alpha)^2}$.
 It is easy to confirm its very weak time dependence, since
 the power-law index is limited to the range 
 $0\leq (2-\alpha)/(4-\alpha)^2 \leq 1/9$.
 Just in the same way, if the constant aspect ratio is adopted (Case
 I), we find again a very weak time dependence of the opening angle, 
 ${\rm tan}\theta(t) \propto t^{(\alpha - 2)/[2(5-\alpha)]}$,
 with the power-law index being $-1/8\leq (\alpha - 2)/[2(5-\alpha)] \leq
 0$.

 As a consequence of this small difference between
 the two solutions, the corresponding values of $X$
 also show only a slight difference.
 For example, when a typical value $\alpha = 1.5$
 is taken, we obtain $X=12/7\sim1.71$ and $X= 1.76$ for Case I  
  and Case II, respectively. 
%
 Hence, 
 the estimated $L_{\rm j}$ and
 $t_{\rm age}$ based on
 these two sets of solutions also do not vary much, either.
 Indeed, for the given values of $r_{\rm h}$,
 $A_{\rm h}$, and ${\cal R}$,
 the derived $L_{\rm j}$ and $t_{\rm age}$
 only differ by a factor of $\sim 1.7$ and $\sim 0.83$, respectively,
 between the two cases for $\alpha = 1.5$.
 It is worth noting that
 for $\alpha = 2$, which corresponds to the
 to 3C 219 (Table \ref{tab1}),
 the two conditions give the same value of $X=2$ and, as a result,
 the solutions are identical.    
 Since 
 only a slight change is found between the two cases,
 we focus on the widely-discussed self-similar
 solution (Case I) in the following.

%



\subsection{Improvements from KK05}
\label{improve}

In the present study,
we have improved the energy equation given in KK05
for more accurate quantitative estimations of 
$L_{\rm j}$  and $t_{\rm age}$.
As mentioned in \S \ref{beq},
the resultant change 
in the derived $L_{\rm j}$
turns out to be rather large.
Here we explain the reasons for this discrepancy
more in detail.

As mentioned in \S \ref{beq}, we have (i) modified $dV_{\rm c}/dt$ and
(ii) included the $PdV$ term in Eq. (\ref{energy}).
 As for (i), the main flaw in KK05 
 is the fact that they did not take into account the 
  sideways expansion in the growth of $V_{\rm c}$. In fact, they
 employed the equation, $dV_{\rm c}/dt = 2 \pi r_{\rm c}^2 v_{\rm h}$, 
 whereas a more accurate expression is 
 $dV_{\rm c}/dt =
 (4/3)[\pi r_{\rm c}^2 v_{\rm h} + 2\pi r_{\rm c} r_{\rm h} v_{\rm c}]$,
 which is adopted in the present study.
 As for (ii),
 it was assumed that all injected energy
 is converted to internal energy 
 (namely, $E_{\rm c} = 2L_{\rm j}t_{\rm age}$) 
 in KK05.
 it is obvious, however, that a part of the injected energy is
 consumed for expansions and the $PdV$ work should be included,
 particularly for quantitative estimations.
 As found in \S \ref{Ana},
 these corrections are reflected only in the numerical factor 
 ${\cal C}_{\rm vh}$ in Eq.~(\ref{v_h}).
 The value of ${\cal C}_{\rm vh}$ is reduced by a factor of $\sim 3.5$
 owing to (i) and another factor of $\sim 2$ due to (ii) and, hence, by
 a factor of $\sim 7$ as a whole.
 For a given geometry of cocoon
 ($r_{\rm h}$, $A_{\rm h}$, and ${\cal R}$) and
 ambient density profile ($\rho_{\rm a}$ and $\alpha$),
 the derived power and age scale with 
 the numerical factor as $L_{\rm j} \propto {\cal C}_{\rm vh}^2$
 and $t_{\rm age} \propto {\cal C}_{\rm vh}^{-1}$, respectively.
 As a result, KK05 overestimated $L_{\rm j}$ by a factor of $\sim 50$.
 On the other hand, $t_{\rm age}$ was underestimated by a factor of 
 $\sim 0.14$, which led to the overestimation of $E_{\rm c}$ 
 by a factor of $\sim 14$, since the latter  is obtained as
 $E_{\rm c}=2L_{\rm j}t_{\rm age}$ in KK05.  
 It is also worthy to note that in the present study the following
 relation holds: $E_{\rm c}\simeq L_{\rm j}t_{\rm age}$. 
 The difference of the factor $\sim 2$ arises from
 the fact that about a half of the injected energy is used for the 
 $PdV$ work.

\section{EXTRACTION OF THE KEY QUANTITIES FROM THE OBSERVATIONS}
\label{extraction}

In determining $L_{\rm j}$ and $t_{\rm age}$, we
essentially follow the same procedure taken in KK05. 
In this section, we explain in detail
 how to extract the key quantities utilized in the model from 
the observations. 

\subsection{ICM quantities}

As for the mass-density profiles ($\rho_{\rm a}$) and
pressures ($P_{\rm a}$) of ICM, 
we adopt the values given in the literature
(\citet{RF96, SW02} for Cygnus A,
\citet{CB04} for 3C 223 and 3C 284,
and \citet{HW99}  for 3C 219). 
 In these papers,  the X-ray surface brightness
 distribution of  ICM  
  was fitted 
by the isothermal  $\beta$-model, 
which takes the form of
 $\rho(r) = \rho_{\rm core} [1 + (r/r_{\rm core})^{2}]^{-3\beta/2}$
 \citep{CF78},
where $\rho_{\rm core}$ and $r_{\rm core}$ are the core radius and 
 density of the ICM, respectively.
 Since we employ the density profile of
 $\rho_{\rm a}(r) = {\bar \rho_{\rm a}} (r/r_{\rm h})^{-\alpha}$
 in our model,
 a power-law approximation
 of the $\beta$-model  is necessary.
 In the present study, we determine
  ${\bar \rho}_{\rm a}=\rho_{\rm a}(r_{\rm h})$ and
  $\alpha$ from the $\beta$-model as follows.
 The determination of ${\bar \rho}_{\rm a}$ is 
 done simply by equating it with the density in the $\beta$-model at
 the corresponding
 radius $r_{\rm h}$,
 namely
 ${\bar \rho_{\rm a}} = \rho_{\rm core}
 [1 + (r_{\rm h} / r_{\rm core})^{2}]^{-3\beta/2}$.  
 In the case of $r_{\rm h}\gg r_{\rm core}$, 
 it is clear that $\alpha$ can be approximated by $3\beta$.
 Only Cygnus A satisfies this condition, though.
 For the rest of the sources, 
 $r_{\rm h}$ is comparable to $r_{\rm core}$:
 $r_{\rm core}$  $\sim$ 340 kpc,  210 kpc, and 90 kpc
 for 3C 223, 3C 284, and 3C 219,
 respectively.
 It is obvious that the above approximation of $\alpha \approx 3\beta$
 would cause an overestimation of density gradient for these cases.
 Instead  $\alpha$ should
be taken to be a typical value in the ICM region 
swept by the expanding  cocoon.
 Here we determine $\alpha$ by requiring
 that $\rho_{\rm a}(r)$ should coincide with the
 density in the  $\beta$-model at  $r = 0.5 r_{\rm h}$
 in addition to $r = r_{\rm h}$. 
Although there is no compelling reason for
the choice of $r = 0.5 r_{\rm h}$,
 the estimations of
 $L_{\rm j}$ and $t_{\rm age}$
 are affected little by this uncertainty (\S\ref{result1}).
 Once  $\rho_{\rm a}$ is given,
$P_{\rm a}$ is  obtained by the 
equation of state, which is written as
$P_a(r) = \frac{k_B T_{\rm a}}{\mu m_{\rm H}} \rho_a(r)$,
 where $T_{\rm a}$ and $\mu=0.6$ are
 the temperature  and
 mean molecular weight of ICM, respectively, and 
 $m_{\rm H}$ is the mass of hydrogen.           
 We adopt the temperature used in the $\beta$-model, ignoring 
the radial dependence of $T_{\rm a}$ as usual.
In  Table \ref{tab1}, we list the  values of
$\rho_{\rm a}$ and
$P_{\rm a}$ at $r=r_{\rm h}$ and
$\alpha$ for each source.

\subsection{$r_{\rm h}$ and $A_{\rm h}$}

In Fig. \ref{radio}, we show the 
VLA images of 
Cygnus A \citep{PDC84},
3C 223 \citep{LP91}, 
3C 284 \citep{LPR86}, and 
3C 219 \citep{CBBP92} in logarithmic scale. 
 Contours in linear scale are also displayed  to
 determine the position of hot spot accurately.
 The overlaid straight lines that cross  each 
 other at right angle on the hot spot are the lines we use to measure 
$r_{\rm h}$ and $A_{\rm h}$. 
For simplicity, we neglect the projection effect of $r_{\rm h}$,
which would be at most
 a factor of a few 
if we believe the unified model of AGN \citep{UP95}.
 $A_{\rm h}$ is measured as a cross-sectional
 area of the radio lobe at the position of the hot spot.
 $r_{\rm h}$ and $A_{\rm h}$ for  each source
 are summarized in Table \ref{tab1}.

 It should be noted that 
 the plasma just around the hot spot is very fresh in the sense that
 a significant synchrotron cooling is absent. 
 Hence,  the effect of radiative cooling does 
 not introduce large ambiguity in the estimation of $A_{\rm h}$.
 The adiabatic cooling, on the other hand, is not expected to
 cause any problem in the estimation for the following reason.
 Since the sound crossing time in the head region of
 the cocoon, $\sim A_{\rm h}^{1/2} / c_{\rm s}$, where  
 $c_{\rm s}=c/\sqrt{3}$ is the sound speed, is much
 shorter than the age of the cocoon, we can regard the
 head region to be uniform to the lowest order.
 Hence, the adiabatic cooling, if any,
 would decrease the surface brightness gradually as the distance from 
 the hot spot increases.
  Contrary to this, the observed radio images
   show a sharp decline of the 
   surface brightness at the outer edge, which is most naturally interpreted 
   as the existence of the periphery of cocoon head there.
%

\begin{figure}[ht]
\begin{center} 
\includegraphics[width=9cm]{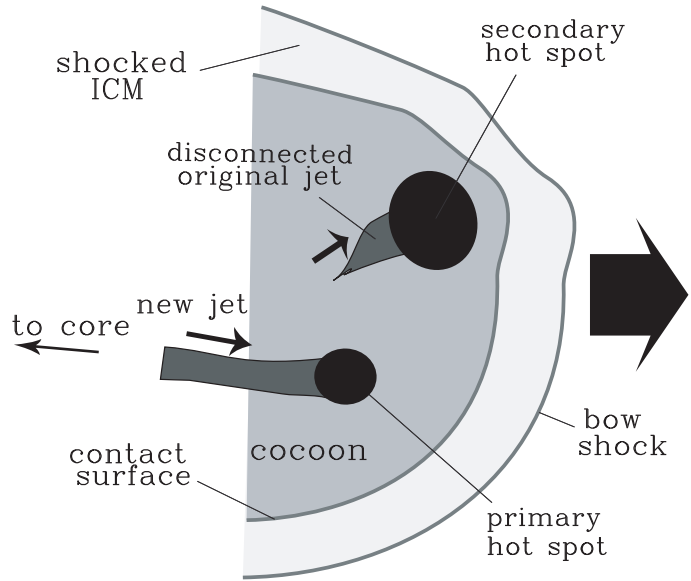}
\caption
{A cartoon around the  double hot spots in the ``disconnected-jet'' model. 
As a result of altering orientation, jet becomes disconnected
and forms the primary and the secondary hot spot which are
 the location of the present and the previous terminal shock, respectively.}
\label{double}
\end{center}
\end{figure}

Next we address the issue in determining $A_{\rm h}$ 
that arises from multiple hot spots.
Double hot spots are actually
observed in the radio lobes of Cygnus A
 (see, e.g., Carilli \& Barthel 1996).
In determining $A_{\rm h}$,  
we adopt the ``disconnected-jet'' model by \citet{CG91}
to Cygnus A.
 The double hot spots are referred to 
as primary and secondary as follows.  
The primary hot spot 
is  more compact and
located in the inner part of the lobe,
whereas  the secondary spot is 
more diffuse and brighter and located  in the outer part
of the lobe. 
According to the disconnected-jet model,
double hot spots are produced by the sudden change of the
jet-orientation,
or the disconnection, 
which leads to the termination of energy supply to the original shock and
the generation of a new jet.
While the primary hot spot is produced by 
the  terminal shock in the new jet, 
the secondary hot spot remains as a relic in the original jet.
The schematic picture of the model is illustrated in Fig. \ref{double}.
Since the primary hot spot is predicted to be
much younger than  
the source age,
 we employ  the position
of the secondary hot spot to determine $A_{\rm h}$.
We will discuss this point more in detail in the following.

 The age of Cygnus A is roughly estimated to be 
 $t_{\rm age}
 \approx r_{\rm h}/\beta_{\rm hs}c
 \approx 2.0\times 10^{7}(\beta_{\rm hs}/10^{-2})^{-1}$ yr,
 where $\beta_{\rm hs}c$ is the advance velocity of the hot spot. 
 On the other hand, 
 since the primary hot spot is observed  simultaneously 
 with the secondary hot spot,
 its age should be younger than the duration,  $t_{\rm dur}$,
 in which the secondary spot is bright.
 $t_{\rm dur}$
is expressed as a sum of the
time up to the shut-off of the energy
 supply from the disconnected-jet to the spot, $t_{\rm dis}$,
and the cooling time, $t_{\rm cool}$:
 \begin{eqnarray} t_{\rm dur} = t_{\rm dis} + t_{\rm cool} \approx
 {\rm max}(t_{\rm dis}, t_{\rm cool}) . \label{duration} 
\end{eqnarray}
The cooling time  is evaluated as
 $t_{\rm cool} = {\rm min}(t_{\rm syn}, t_{\rm ad})$,
 where  $t_{\rm syn}$ and $t_{\rm ad}$ are
 the synchrotron cooling timescale and the adiabatic expansion
 timescale, respectively.
A typical value of  $t_{\rm syn}$ at the hot spot is 
estimated as  $t_{\rm syn} \approx 1.0 \times 10^6 
                 ({B}/{10^{-4}{\rm G}})^{-3/2} 
                 ( {\nu}/{1~{\rm GHz}})^{-1/3}
                  ~{\rm yr}$,
 whilst $t_{\rm ad}$ is given by
 $t_{\rm ad} \approx r_{\rm hs} / c_{\rm s}
 = 5.6\times10^{3}(r_{\rm hs}/1{\rm kpc})~{\rm yr}$, 
 where $r_{\rm hs}$
 is the size of the hot spot.
 These estimates lead to $t_{\rm cool} = t_{\rm ad}$  
for the secondary hot spot.
 On the other hand,  
$t_{\rm dis}$ is given by
 $t_{\rm dis}=r_{\rm dis} / v_{\rm j}$,    
 where
 $r_{\rm dis}$ is the distance from the tail to the hot spot in the
 disconnected jet. 
  Although we do not know  $r_{\rm dis}$ from observations,
  we can at least put  the upper limit as 
  $r_{\rm h}>r_{\rm dis}$.
We then obtain $t_{\rm dis} < 2.0 \times 10^{5}
  (r_{\rm h}/60~{\rm kpc})(v_{\rm j}/c)^{-1}$ yr.
 Hence, from Eq. (\ref{duration}), we see that  $t_{\rm dur}$ is in the range
 $\sim 5\times 10^3 \-- 2\times10^5$ yr.
 From  these estimates, it is obtained that 
 the age of the primary hot spot 
 only makes up a small fraction of its whole lifetime. 
 Therefore,  we adopt the
 secondary spot for the determination of $A_{\rm h}$, which 
reflects the whole evolution  of the cocoon (Fig. \ref{radio}).

\subsection{$\cal{R}$}
\label{calR}

 In contrast to $r_{\rm h}$ and $A_{\rm h}$,
 it is difficult to measure the aspect ratio of cocoon, $\cal{R}$, 
from the VLA radio images,
since the  cocoon emission  from 
the region far away from the hot spot
 is very dim at GHz  frequency 
because of the synchrotron cooling (see Fig. \ref{radio}).
It is well known that  this cooling effect can be used to infer the age 
of radio galaxy
 (the spectral ageing method, e.g., Carilli et al. 1991).
As one utilizes lower frequencies, however, the cocoon 
image is expected to be thicker,
since lower-energy electrons
have longer synchrotron cooling times
(e.g., Carilli and Barthel 1996). 
In fact, a few authors have reported 
the existence of  prolate cocoons,
based on the observations 
 at a relatively low radio frequency (610 MHz) band 
(e.g., Readhead et al. 1996).
It is mentioned, however, that little attention has been paid to
observational features concerning the structure of cocoons so far.


On the other hand, theoretical
studies of jet propagation and cocoon formation with multi-dimensional
hydrodynamic simulations clearly support the existence of  cocoon for
reasonably light beams going through surrounding ICM
\citep{SA02, AG00, GM97, KF97, MHD97}. 
The intensity maps of the synchrotron emissivities obtained
in these hydrodynamical simulations  
well reproduce the
double lobe structures as observed (e.g., Fig. 10 in Scheck et al. 2002).
It seems natural, therefore, to suppose that 
cocoons are commonly produced, although there is 
a room for further observational investigations.
In the present study, we explore a  wide range of $0.5<{\cal R}<1$
in order to take account of the large ambiguity on the shape of cocoon.
 It is worthwhile to note in this respect that the 
 existence of cocoon can be confirmed for Cygnus A in the Chandra image
 \citep{WY00, WSY06} and that the obtained aspect ratio $\sim 0.5 - 0.7$ 
 lies in the range explored in this paper.

\section{RESULTS}

\subsection{Total kinetic power and dynamical age}
\label{result1}

The resultant $L_{\rm j}$ and $t_{\rm age}$ 
are displayed in Figs. \ref{CygApower}, \ref{3C223power},
\ref{3C284power}, and \ref{3C219power}.
 Since
 most of the
 radio sources show asymmetries in  the pair 
 of  lobes,
 we analyze each  lobe independently. 
 Three oblique lines  in these figures are 
 the obtained $L_{\rm j}$ and $t_{\rm age}$ 
for different $\alpha$'s, and their ranges reflect
 the uncertainty in $\cal{R}$. 
The thick solid line shows the result for 
the parameter set  listed in Table \ref{tab1}.
 The other two lines correspond to the results obtained 
 by varying $\alpha$ by $\pm$0.5.
 From the figures, it is confirmed that the results are
 rather insensitive to the value $\alpha$. 
 In each line, the power and age depend on the aspect ratio $\cal{R}$ as 
 $L_{\rm j} \propto \cal{R}$$^{ 2 \alpha - 8}$  and
 $t_{\rm age} \propto \cal{R}$$^{ 4 - \alpha}$, 
 and, therefore,  satisfy 
 $L_{\rm j} \propto t_{\rm age}^{ -2 }$.
 Since $\alpha$  does not exceed $4$ in any of the four sources,
 a lower aspect ratio corresponds to
 a higher power with a lower age.
It should be noted that the  
 uncertainty in the absolute values of $\rho_{\rm a}$ and $P_{\rm a}$ 
 is not crucial,
 since
$P_{\rm a}$ is used only to judge 
 whether the over-pressure condition is 
satisfied or not,
 and the dependence of $L_{\rm j}$ and $t_{\rm age}$ on
 $\rho_{\rm a}$ is rather weak,
 $L_{\rm j} \propto \rho_{\rm a}$ and 
 $t_{\rm age} \propto \rho_{\rm a}^0$. 
 The line outside the shaded region must be discarded,
 since the over-pressure condition is violated. 
 The Eddington luminosity, $L_{\rm Edd}$, of each source 
 is also shown in these figures for comparison.
In Table \ref{tab3}, we summarize 
the allowed values of $L_{\rm j}$ and $t_{\rm age}$
 and  other relevant physical parameters of  cocoon
 obtained for the parameter set listed in Table \ref{tab1}.

 Cygnus A is one of the vastly studied nearby
 FR II radio galaxies.
 \citet{TMA03} estimated the SMBH mass of Cygnus A as
 $2.5\times 10^9  M_{\odot}$, based on the gas kinematics in the narrow-line
 region.
 Its linear size is measured as $r_{\rm h} = 70$ kpc for the western jet
 and  $r_{\rm h} = 60$ kpc for the eastern jet.
 From the employed values of $r_{\rm h}$ and $A_{\rm h}$,
 the power and age are obtained as
 $L_{\rm j} = (0.35 \-- 1.1)\times 10^{46}~{\rm ergs~s^{-1}}$ and
 $t_{\rm age} = 30 - 53~{\rm Myr}$ for the western jet 
 and 
 $L_{\rm j} = (0.4 - 2.6)\times 10^{46}~{\rm ergs~s^{-1}}$ and
 $t_{\rm age} = 19 - 47~{\rm Myr}$ for the eastern jet.
 No significant difference is seen between the two jets, and
 we interpret that the actual age lies in these ranges.
Note that
  $L_{\rm j}$  is  decreased and $t_{\rm age}$ is 
 increased from those in KK05 by the improvement of $V_{\rm c}$
 and the inclusion of $PdV$ work.
 In fact,  while $L_{\rm j} = 1.3\times 10^{48}~{\rm ergs~s}^{-1}$ and
 $t_{\rm age} = 2.6~{\rm Myr}$ were obtained in KK05, we find
 $L_{\rm j} = 2.6\times 10^{46}~{\rm ergs~s}^{-1}$ and
 $t_{\rm age} = 19~{\rm Myr}$  
 in this paper when the identical values of
 $r_{\rm h} = 60$ kpc, $A_{\rm h} = 150~{\rm kpc}^{2}$,
 and $\cal{R} = $ 0.5 are employed.

\begin{figure}
 \begin{center}
\includegraphics[angle=0,scale=.50]{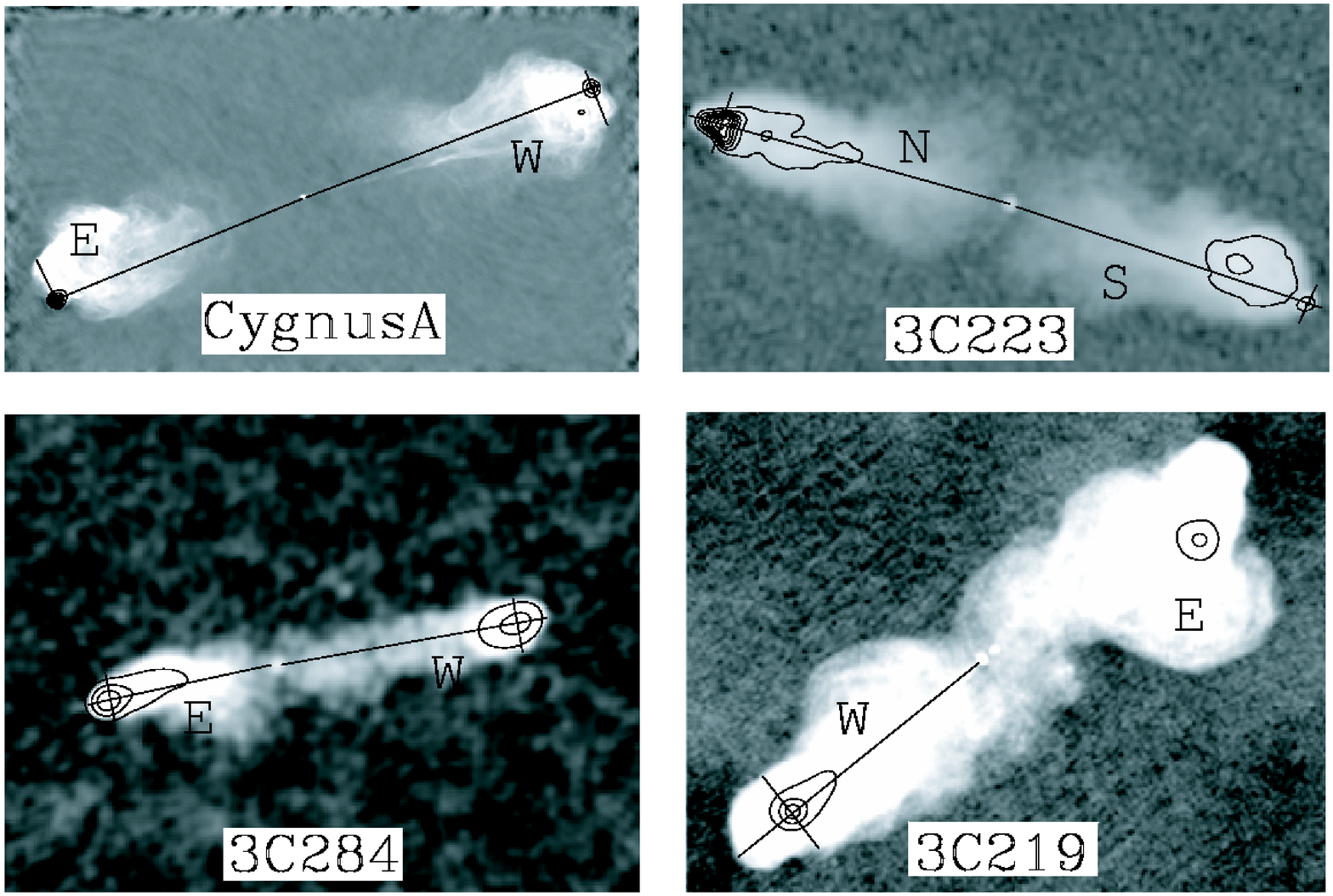}
\caption{Logarithmic-scaled 5-GHz VLA map of Cygnus A (upper-left) and 1.5GHz VLA maps of
  3C 223 (upper-right), 3C 284 (lower-left), and
 3C 219 (lower-right) with linearly spaced contours are displayed.  
 The straight 
 lines overlaid in each map denote the lines we have used
 to measure $r_{\rm h}$ and $A_{\rm h}$.}
\label{radio}
 \end{center}
\end{figure}

 3C 223 has  radio lobes that  extend up to
 $r_{\rm h} = 340$ kpc in both sides.   
 Its SMBH mass is estimated as $1.4 \times 10^{8}M_{\odot}$
 by \citet{WU02}, based on the observed stellar velocity
 dispersions.
 As can be seen in Fig. \ref{radio},
 3C 223 has asymmetry in $A_{\rm h}$.
 While a well developed cocoon head is seen 
 at the northern hot spot ($A_{\rm h} = 4300~{\rm kpc}^2$), 
 the cocoon head 
 at the southern hot spot 
 is quite compact  ($A_{\rm h} = 1800~{\rm kpc}^2$).
 Reflecting this asymmetry,
 the obtained power and age
 show quite large difference:
 $L_{\rm j} = (0.15 - 2.9)\times 10^{46}~{\rm ergs~s}^{-1}$
 and $t_{\rm age} = 140 - 610$ Myr for the northern jet, 
 and  $L_{\rm j} = (0.71 - 2.0)\times 10^{45}~{\rm ergs~s}^{-1}$
 and $t_{\rm age} = 330 - 560$ Myr for the southern jet.

 3C 284  shows  
 asymmetry both in $r_{\rm h}$ and $A_{\rm h}$.
 While $r_{\rm h}$ and $A_{\rm h}$
 in the western lobe are estimated as $ 380$ kpc and
 $ 6200$ kpc$^2$, respectively, 
 the corresponding values for the eastern lobe are $ 260$ kpc
 and  $4600$ kpc$^2$.
 The obtained power and age
 are
 $L_{\rm j} = (0.1 - 3.6)\times 10^{46}~{\rm ergs~s}^{-1}$ and
 $t_{\rm age} = 100 - 630$ Myr for the western jet and 
 $L_{\rm j} = (0.03 - 1.8)\times 10^{47}~{\rm ergs~s}^{-1}$ and  
 $t_{\rm age} = 32 - 260$ Myr for the eastern jet.
  Since there is no  estimate of the
 SMBH mass of 3C 284 in the literature, 
 we derive the mass from the 
 B-band magnitude of the buldge estimated in
 \citet{SRH05}. By using the equation in \citet{MCF04}
 which gives the correlation of the B-band magnitude with the BH mass,
 we obtain $M_{\rm BH} = 8.2\times 10^{8} M_{\odot}$.

 In the case of 3C 219,
 we only analyze the jet on the western side,
 since the eastern lobe shows severe deformation (see
 Fig. \ref{radio}).
  We could not determine $A_{\rm h}$ on the eastern
 side from its morphology.
 The central SMBH mass for 3C 219 is estimated by 
 \citet{MCF04} 
 as $6.3\times 10^{8}~M_{\odot}$.
 $r_{\rm h}$ and $A_{\rm h}$ of the western lobe are
 measured as $ 210$ kpc and $ 5000$ kpc$^2$.
 From these values, the  kinetic power and  age
 are obtained as 
 $L_{\rm j} = (0.26 - 4.3)\times 10^{47}~{\rm ergs~s}^{-1}$ and
 $t_{\rm age} = 37 - 150$ Myr, respectively.

Large asymmetry between the pair of lobes is observed   
especially in 3C 223 and 3C 284, and 3C 219.
Since it seems natural to suppose that
 the jet properties  are intrinsically
symmetric and the power and age are identical on
both sides, 
we expect that  the asymmetry in the pair of lobes is 
due to the asymmetry and/or inhomogeneity in the ICM density profiles. 
Although this is an interesting subject, a further pursuit is beyond the 
scope of the present study.
Here we assume that the actual values of $L_{\rm j}$ and $t_{\rm age}$
are lying 
in the ranges obtained from both lobes.

In Table \ref{tab3} (column 4), the total kinetic powers of  jet
 normalized by the corresponding Eddington luminosity, 
 $2 L_{\rm j} / L_{\rm Edd}$, are
displayed. 
It can be seen that $2 L_{\rm j} / L_{\rm Edd}$ takes 
quite high values ranging from $\sim 0.02$ to $\sim 10$. 
In exploring the physical relations between 
the accretion power and the outflow, 
the total kinetic power of jet normalized by the
Eddington luminosity, $2 L_{\rm j} / L_{\rm Edd}$,
is one of the most fundamental parameters. 
We will return to this issue in \S5.1.

\subsection{Total internal energy vs. minimum energy}
\label{resultmin}

It is intriguing to compare 
the  internal energy, $E_{\rm c}$, 
 deposited in the cocoon with the widely discussed 
 energy, 
$E_{\rm min}$, obtained from the minimum energy condition
\citep[e.g.,][]{M80}.
 $E_{\rm c}$ is linearly proportional to the total 
 energy injected in the cocoon ($2L_{\rm j}t_{\rm age}$)
 and is approximately given as $E_{\rm c} \simeq L_{\rm j}t_{\rm age}$.
 The dependence on the aspect ratio is given by
 $E_{\rm c}\propto L_{\rm j}t_{\rm age}\propto L_{\rm j}^{1/2}
 \propto {\cal R}^{\alpha - 4}$
 (\S \ref{result1}).
 Hence, a smaller  aspect ratio
 (or, equivalently, a larger power) corresponds to
 a larger internal energy.
 $E_{\rm min}$ is the minimum value of the
 total energy (the sum of the energies in radiating non-thermal electrons
 and magnetic fields) required for a
 given synchrotron luminosity and is evaluated as
\begin{eqnarray}
 E_{\rm min}
 = 
\frac{7}{24 \pi} V_{\rm R}^{3/7} 
  \Bigl[12\pi^{1/2}  f(\alpha_{\rm R}) (\nu_{\rm min}^{0.5 - \alpha_{\rm R}} 
     -  \nu_{\rm max}^{0.5 - \alpha_{\rm R}}) \nu^{\alpha_{\rm R}}
              L_{\nu} \Bigr]^{4/7}~~{\rm ergs},    
\label{minimum}
\end{eqnarray}
 where $V_{\rm R}$ is the volume of the emitting region,
 $\alpha_{\rm R}$ is the spectral index of the synchrotron emission,  
 and 
$L_{\nu}$ 
is the synchrotron luminosity  measured at frequency $\nu$,
 and  $\nu_{\rm min}$ and $\nu_{\rm max}$ are the lower  and
 higher cut-offs in the synchrotron emission, 
 respectively.
 $f(\alpha_{\rm R})$ is a function of spectral index
 $\alpha_{\rm R}$ which
 is  given as
\begin{eqnarray}
 f(\alpha_{\rm R}) \simeq
 \frac{3.16 \times 10^{12}(0.145)^{\alpha_{\rm R}}
        (2\alpha_{\rm R} + 1) \Gamma(\frac{\alpha_{\rm R}}{2} + 1)}
      {(2\alpha_{\rm R}-1)
       \Gamma(\frac{\alpha_{\rm R}}{2} + \frac{11}{6})
       \Gamma(\frac{\alpha_{\rm R}}{2} + \frac{1}{6})
       \Gamma(\frac{\alpha_{\rm R}}{2} + \frac{3}{2})
   },\nonumber
\end{eqnarray}
 where $\Gamma$ is the Gamma function (see, e.g., Longair 1994). 

The values of the spectral
 index $\alpha_{\rm R}$ at the low frequency band (178--750~MHz) and
 the flux density, $F_{\nu}$, at 178MHz  are taken from 
 Table 1 in \citet{HAP98}.
  From the employed values of $F_{\nu}$, the synchrotron luminosities are
 calculated  as $L_{\nu}=4 \pi d_{\rm L}^2F_{\nu}$, where
 $d_{\rm L}$ is the luminosity distance.
Although 
 $L_{\nu}$
is the sum of the luminosity  from lobes, jets and hot spots,
no significant overestimate of $L_{\nu}$ is 
expected because the radio emission is
 dominated by the lobe-component for
 most of FRII sources
 \citep{BHL94,HAP98}. 
 The employed values of $\alpha_{\rm R}$, $L_{\nu}$
 and  other relevant quantities are summarized in Table \ref{tab2}.
Here we neglect
the second term in Eq. (\ref{minimum}) 
 in deriving $E_{\rm min}$,  since
$\alpha_{\rm R}>0.5$ is satisfied in all sources.
 The lower cut-off frequency $\nu_{\rm min}$ is taken
as $10^{4}$~Hz. We will comment on this value in the next paragraph.
%
 As noted in \S\ref{calR},
 although GHz radio images do not show 
 a cocoon-shape clearly and only a pair of lobes can be seen,
it is known that radio images at lower frequencies reflect the cocoon
 shape more closely \citep[e.g.,][]{RTX96, CPD91}
 because of the absence of efficient radiative coolings.
%
 Since we utilize a relatively
 low frequency (178MHz) band,
 the volume of the emission region can be
 approximated as $V_{\rm R} \sim V_{\rm c}$.
 Here we employ the median value of ${\cal R}$, 
  i.e. ${\cal R}=0.75$, in evaluating $V_{\rm R}$.
 We define the 
 ratio of  $E_{\rm c}$ to  $E_{\rm min}$ 
as $\eta_{\rm c}$:
\begin{eqnarray}
\eta_{\rm c}
\equiv \frac{E_{\rm c}}{E_{\rm min}} .
\label{ratio}
\end{eqnarray}
%
 The obtained values of $E_{\rm min}$, 
  $E_{\rm c}$, and $\eta_{\rm c}$
 are summarized in Table \ref{tab3}.
 We find that $E_{\rm c}$ is larger than $E_{\rm min}$ and 
  $\eta_{\rm c}$ is in
 the range of 
 $ 4< \eta_{\rm c} <310$.
This implies that there is a substantial deviation
 from the minimum energy condition.
 We will discuss  this topic  more in detail 
 in \S\ref{content}.

Lastly, we comment on the reliability of $\eta_{\rm c}$. 
The lower cut-off frequency,
 $\nu_{\rm min}$, is one of the ingredients, which introduce 
uncertainties in $\eta_{\rm c}$
because it is difficult to determine by radio observations.
 The value employed above is  obtained from the following relation
 $\nu_{\rm min} \approx10^{4}~(B/10^{-5}~{\rm G})
  (\gamma_{\rm e,min}/10)^2~{\rm Hz}$,
 where
 $\gamma_{\rm e,min}$ is the minimum Lorentz factor of non-thermal
 electrons.
 Note that the
 the resultant $E_{\rm min}$ does not change significantly by
 the uncertainty in $\nu_{\rm min}$  because of its weak dependence.
 For example, when a typical value of  $\alpha_{\rm R} = 0.8$
 is employed 
$ E_{\rm min} \propto
 (\nu_{\rm min} / 10^4~{\rm Hz})^{6/35}$. 
It should be also mentioned that  the 
 difference between the actual emission
 volume and the employed one, which we 
 do not expect to vary by orders,
 does not
 affect our result, since the dependence 
 of $E_{\rm min}$ on the emission volume $V_{\rm R}$ is weak, 
$E_{\rm min}\propto V_{\rm R}^{3/7}$. Hence the precise determination of
 the latter is not necessary.


\subsection{On the estimation of $L_{\rm j}$ and $E_{\rm c}$}
\label{estimate}

\subsubsection{Upper limits and lower limits}

\begin{figure}[ht]
\begin{center} 
\includegraphics[width=14cm]{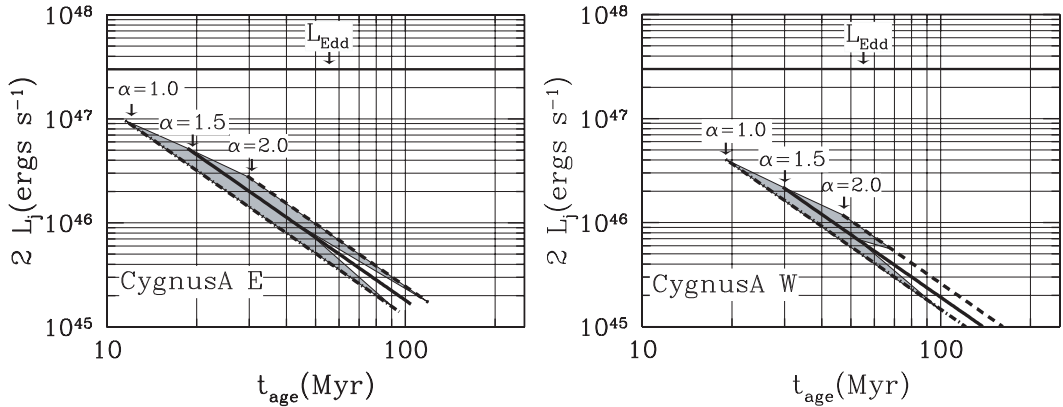}
\caption
{The obtained ranges of power and age of Cygnus A.
 The three oblique lines, which lie closely to each other
 are the solutions for 
the power-law index of the ICM density ($\alpha$) shown by
 arrow;
 The solid line represents the solution for the estimated power-law
 index (see Table \ref{tab1}). The dashed  and 
 dot-dashed lines represent the solutions for the power-law index
 increased by $0.5$ and decreased by $0.5$, respectively.
 The shaded regions show  allowed ranges where the overpressure
 condition ($P_{\rm c} > P_{\rm a}$) is satisfied.
 Also  the Eddington luminosities are displayed 
 by the horizontal lines for comparison.} 
\label{CygApower}
\end{center}
\end{figure}

\begin{figure}[ht]
\begin{center} 
\includegraphics[width=14cm]{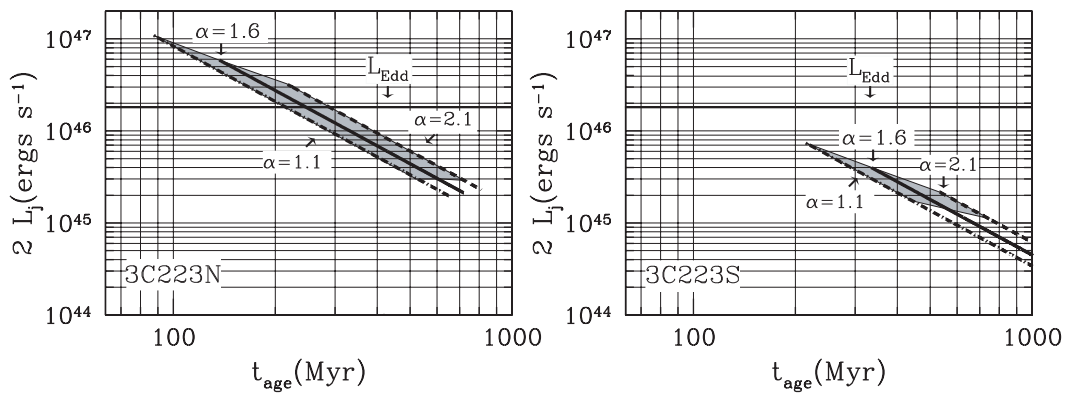}
\caption
{Same as Fig. \ref{CygApower}, but for 3C 223.}
\label{3C223power}
\end{center}
\end{figure}

\begin{figure}[ht]
\begin{center} 
\includegraphics[width=14cm]{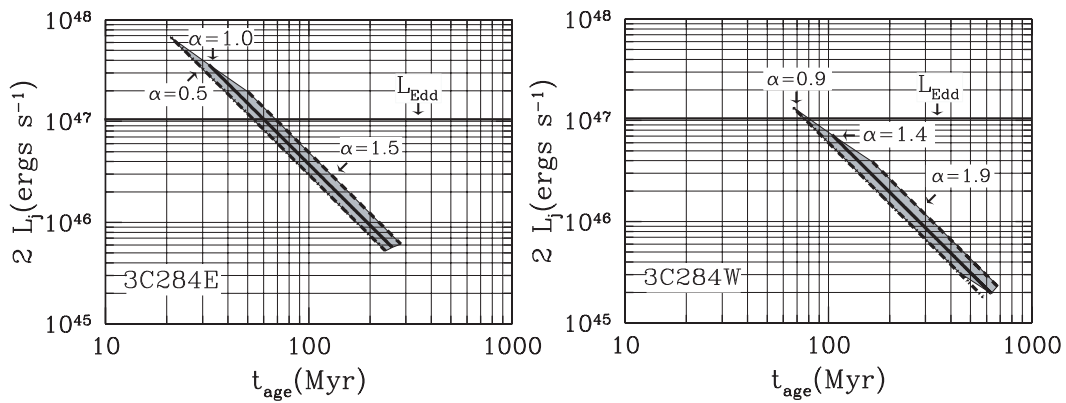}
\caption
{Same as Fig. \ref{CygApower}, but for 3C 284.}
\label{3C284power}
\end{center}
\end{figure}

\begin{figure}[ht]
\begin{center} 
\includegraphics[width=7cm]{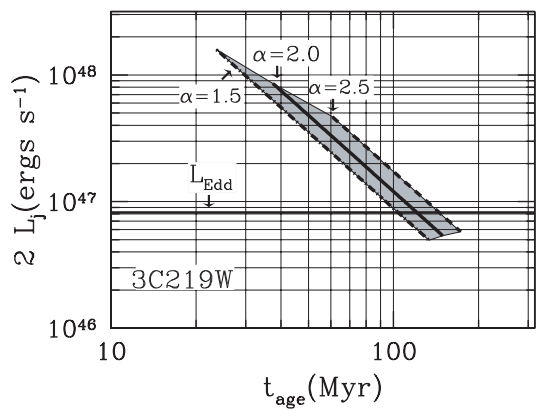}
\caption
{Same as Fig. \ref{CygApower}, but for 3C 219.}
\label{3C219power}
\end{center}
\end{figure}

 It is important to consider the validity of the over-pressure
 condition  (i.e. $P_{\rm c}> P_{\rm a}$), since
 the lower limits of
 $L_{\rm j}$ and $E_{\rm c}$ are
 determined by this condition  in most cases (see
 Figs. \ref{CygApower}-\ref{3C219power}).
 In our  model, a larger $P_{\rm c}$ corresponds to a smaller 
${\cal R}$. 
 Though we explore a wide range of ${\cal R}$ ($0.5 \sim 1$), 
it is intuitively more likely that ${\cal R}$ 
 is smaller than unity, that is, the cocoon is prolate rather
 than spherical.
 Hence, the 
 results of our analysis suggest that the sources examined
 in the present study are likely to be over-pressured indeed.
  Incidentally, the prolate shape of cocoon is endorsed by the
 fact that independent age estimations of Cygnus A
 \citep{CPD91} and 3C 284 \citep{AL87} based on the 
 spectral ageing method are more consistent with the results
 for the aspect ratio of $0.5$ than for $1.0$.
 (Unfortunately, the age estimations are not available for 3C 223
 and 3C 219 in the literature).

 In all cases,
 the maximum values of $L_{\rm j}$
 and $E_{\rm c}$ correspond to the minimum value of ${\cal R}$ 
 ($L_{\rm j} \propto {\cal{R}}^{2 \alpha - 8}$).
 Though  ${\cal R}=0.5$ is chosen as the lower limit
 in the present study (\S\ref{calR}),
 the possibility of even smaller aspect ratios cannot
 be ruled out, since the radius of cocoon body
 $r_{\rm c}$ cannot be constrained very well from the radio images.
 It is emphasized again 
 that smaller values of ${\cal R}$ predict 
 larger $L_{\rm j}$ 
 and  $E_{\rm c}$.
 

\subsubsection{Kinetic energy in the cocoon}

We have so far neglected the kinetic energy of bulk flows in the
cocoon, assuming that the internal energy is dominated over the
kinetic energy. This may be justified by the fact that the cocoon is
filled with shocked jet matter, which are expected to flow
subsonically. It is, 
however, important to estimate the possible changes in $L_{\rm j}$ and
$E_{\rm c}$ that the inclusion of the kinetic energy in 
Eq. (\ref{energy}) (energy equation) may make, 
since our results are rather sensitive to the 
changes in the energy equation (see the discussion in \S \ref{improve}).
Although the lack of our knowledge on the mass deposited in the cocoon makes 
it difficult to estimate the kinetic energy quantitatively, 
the changes in $L_{\rm j}$ and 
$E_{\rm c}$ are not significant for the conclusion of the paper 
even in the case, where the kinetic energy is comparable
to the internal energy, as shown shortly.

Just as in \S \ref{improve}, the modification in energy equation is 
reflected in the value of the numerical factor ${\cal C}_{\rm vh}$ of
Eq.~(\ref{v_h}).
%
%
Note that ${\cal C}_{\rm vh}$ depends linearly on the ratio of the 
total internal energy to the total injected energy, 
$\epsilon \equiv E_{\rm c}/(2L_{\rm j}t_{\rm age})$, 
which is $\sim 1/2$ in the present study, since roughly a 
half of the injected energy is consumed for the $PdV$ work.  
The inclusion of the kinetic energy modifies 
$\epsilon$ as $\epsilon \sim 1/(2+f)$, where $f$ is defined 
as the ratio of the kinetic energy to the internal energy and 
${\cal C}_{\rm vh}$ is reduced from the value obtained in the
present study by a factor of $\sim 2/(2+f)$.
It is hence found that even if the kinetic energy is as
large as the internal energy, namely $f\sim 1$,
$L_{\rm j} (\propto {\cal C}_{\rm vh}^2)$  and
 $E_{\rm c} (\propto {\cal C}_{\rm vh})$ are reduced only by 
factors of $\sim 4/9$ and $\sim 2/3$, respectively.
The inclusion of the kinetic energy, therefore, does not change the
conclusion of this paper. 

\section{SUMMARY AND DISCUSSIONS}
\label{SD}

In this paper we have investigated the 
total kinetic power and age of  powerful FR II jets.
We have selected four FR II radio galaxies
(Cygnus A, 3C 223, 3C 284, and 3C 219),
for which the surrounding ICM densities and pressures
are known in the literature.
Below we summarize our main results.

(I) {\it  Large fractions of the Eddington power 
in the range of  $\gtrsim 0.02 - 0.7$ are carried away 
as a kinetic power of  jet in the FR II sources.}

(II) {\it The energy deposited in the cocoon, $E_{\rm c}$,
 exceeds the minimum energy, $E_{\rm min}$,
 by a factor of 
 $4 - 310$.} 

 Although our results  allow a wide range of $L_{\rm j}$
 and $E_{\rm c}$, interesting implications can still be obtained
%
and will be discussed below.
In \S 5.1., we address some issues concerning the ratio of $L_{\rm j}$
to $L_{\rm Edd}$ by referring to the studies of X-ray binaries.
In \S 5.2., the energetics in the cocoon is constrained 
from the obtained $\eta_{\rm c}$.

\subsection{$L_{\rm j} / L_{\rm Edd}$}

Postulating that the relativistic jet is 
 powered by the release  of 
gravitational energy ($L_{\rm acc}$) of accreting matter
(e.g., Marscher et al. 2002), 
 the jet power can be expressed as
 $2L_{\rm j} = \epsilon_{\rm j} L_{\rm acc}$, where
 $\epsilon_{\rm j}$ is the efficiency of energy conversion
 ($0<\epsilon_{\rm j}<1$).
 Hence, $2 L_{\rm j} / L_{\rm Edd}$ 
 gives the minimum mass accretion rate
normalized by the Eddington mass accretion rate.
Our results then suggest that quite high
   mass accretion rates, at least above $0.02L_{\rm Edd}$, are
  required to produce FRII radio sources.
  Moreover, since $10 \gtrsim 2 L_{\rm j} /L_{\rm Edd} \gtrsim 0.65$
  is predicted for 3C 219, some FRII radio sources may have
 super-Eddington mass accretion rates.
The theory of accretion disk predicts that
the accretion disk of these objects is optically-thick and called the {\it
slim-disk} (e.g., Abramowicz et al. 1988). 
From three distinctive X-ray properties (the large photon index 
$\Gamma \gtrsim 2$, rapid variability and soft X-ray excess), 
narrow line Seyfert 1 galaxies (NLS1s) are considered to be
super-Eddington objects and, therefore, are
inferred to have a slim-disk  (e.g., Pounds et al. 1995; Boller et al. 1996; 
Mineshige et al. 2000; Collin \& Kawaguchi 2004; Shemmer et al. 2006). 
If  3C 219 is indeed super-Eddington, it is expected 
to show the above mentioned X-ray features like NLS1s. 
Note, however, that 3C 219 has a relatively hard X-ray spectrum 
with $\Gamma =1.58 <
2 $ (Shi et al. 2005), and thus the physical state of the 
accretion disk in 3C 219 could be different 
from those in NLS1s.

In order to explore the nature of the accretion disk in 3C 219, 
we  compare the characteristics of AGNs with
 those of X-ray binaries (XRB),
since both of them have common physical processes such as disk accretions, 
relativistic jets, and quenching of these jets (e.g., Heinz \& Sunyaev 2003;
Ho 2005; McHardy et al. 2006). 
Thanks to their much shorter dynamical timescales, 
XRBs in various states have been observed in great detail and 
 are found to occupy particular X-ray spectral states 
(Fender et al. 2004; Remillard \& McClintock 2006 for a review)
 as follows;

(i) $L_{\rm acc}/L_{\rm Edd}\lesssim 0.01$ (low/hard state: LS),

(ii) $0.01 \lesssim L_{\rm acc}/L_{\rm Edd}\lesssim 0.3$ 
     (high/soft state: HS),

(iii) $L_{\rm acc}/L_{\rm Edd}\gtrsim 0.3$ (very high state: VHS).

State (i) is usually accompanied by a jet. 
For $L_{\rm acc}/L_{\rm Edd}>0.01$,
the  radio emission is quenched in state (ii), while
in state (iii), the soft VHS, which has an  X-ray spectrum dominated by
a  steep power-law component
 (photon index  $\Gamma >2$), is radio quiet and the 
hard VHS and/or the transition from the hard VHS to the soft VHS 
is accompanied by relativistic ejection events.

If the analogy between XRBs and AGNs holds, NLS1s may be in the soft VHS, 
since many NLS1s have a  steep power-law component and  high
Eddington ratios $>$ 0.3 (e.g., Collin \& Kawaguchi 2004), that is,
higher than the upper limit set by the stability of the 
Shakura \& Sunyaev (1973) disk
 (Shakura \& Sunyaev 1973). 
On the other hand, 3C 219 (FR II) may correspond to the hard VHS 
and/or the transition state because it has 
hard X-ray spectra and high Eddington
luminosities. 
In order to judge whether AGNs  are 
 scaled-up XRBs, it is essential 
to confirm that radio loud AGNs actually have the states analogous to the
 spectral states
(especially VHS) in XRBs\footnote{It has been well established 
that low-luminosity AGNs
are the high-mass counterpart of XRBs in LS's (Ho 2005).}.

The co-evolution of a central black hole (BH) 
and its host galaxy together with AGN feedbacks
have been intensively studied in  various ways 
(e.g., 
Silk \& Rees 1998;
Di Matteo et al. 2003; 
Granato et al. 2004). 
The  intensive surveys of QSOs 
show that the number density of QSOs  is 
peaked at  $z\approx 2$ \citep{FNL01}.
The existence of QSOs at $z\gtrsim 2$  with a central BH of 
a mass smaller than predicted from the bulge BH-mass relation 
 is naturally expected
in the  build-up of SMBHs toward  $z\approx 2$.
For the exploration of the co-evolution processes, dusty spheroidal
galaxies (e.g., galaxies emitting
 sub-millimeter radiations and  ultra-luminous
infrared galaxies) are  the ones to be scrutinized,
since the dusty-gas in them is one of the key quantities 
for the co-evolution. 
Kawakatu et al. (2003)
pointed out the possibility of radio loud AGNs at high-$z$ being
QSOs in the early evolution phase (we call them proto-QSOs) 
that contain a growing BH.
It is, however, difficult to explore the nature of proto-QSOs 
at high-$z$ by  observations in optical and X-ray bands
owing to severe dust-absorptions. 
In contrast, the estimate of 
$L_{\rm j}$ and $t_{\rm age}$ presented in this work 
is applicable even to high-$z$ sources (e.g., Schmidt et al. 2006)
 without being suffered from the dust extinction. 
Therefore, the estimate of 
$L_{\rm j}/L_{\rm Edd}$ based on the cocoon
dynamics is a potential new powerful tool to discover proto-QSOs
among high-$z$ radio loud AGNs.

\subsection{The energetics}
\label{content}

 Lastly, we discuss the energetics in the cocoon.
 Summing up in advance,
 the total internal energy of invisible components such as thermal
 leptons and/or protons tends to be larger than those of radiating non-thermal
 electrons and magnetic fields. 
  $E_{\rm c}$ is expressed by components as
 $E_{\rm c} = (U_{\rm e} + U_{\rm B} + U_{\rm inv})V_{\rm c}$,
 where $U_{\rm e}$, $U_{\rm B}$, and
 $U_{\rm inv}$ are the energy densities
 of  non-thermal leptons (electrons and positrons), magnetic fields,
 and  invisible particles, 
 respectively.
 The energy ratio $\eta_{\rm c}\equiv E_{\rm c}/E_{\rm min}$
 can be then expressed as
 $\eta_{\rm c} =  (U_{\rm e} + U_{\rm B} + U_{\rm inv}) / U_{\rm min}$,
 where  $U_{\rm min}=E_{\rm min}/V_{\rm c}$ is the minimum energy density.
 From the obtained values of $\eta_{\rm c}$,
 we investigate here the contribution of $U_{\rm inv}$
 to the total energy by evaluating $U_{\rm e}/U_{\rm min}$
 and $U_{\rm B}/U_{\rm min}$.

 It is useful to express $U_{\rm e}/U_{\rm min}$ and
 $U_{\rm B}/U_{\rm min}$
 in terms of $U_{\rm e}/U_{\rm B}$, since 
 $U_{\rm e}/U_{\rm B}$ has been intensively investigated by a lot of authors
 (e.g., Isobe et al. 2002; Kataoka et al. 2003;
  Croston et al. 2004; Kataoka \& Stawarz 2005; Croston et al. 2005).
 Since the synchrotron luminosity $L_{\nu}$ is proportional
 to $U_{\rm e}U_{\rm B}^{3/4}V_{\rm c}$,
 we obtain the relation $U_{\rm e}\propto (U_{\rm e}/U_{\rm B})^{3/7}$,
 or equivalently  
 $U_{\rm B}\propto (U_{\rm e}/U_{\rm B})^{-4/7}$ for
 fixed values of $L_{\nu}$ and $V_{\rm c}$.
 From this relation and Eq. (\ref{minimum}), we can
 derive the  following expressions:
\begin{eqnarray}
\frac{U_{\rm e}}{U_{\rm min}} 
\simeq 0.5 \left(\frac{U_{\rm e}}{U_{\rm B}}\right)^{3/7},
 \qquad
 \frac{U_{\rm B}}{U_{\rm min}} 
\simeq 0.5 \left(\frac{U_{\rm e}}{U_{\rm B}}\right)^{-4/7}.
\label{ratio2}
\end{eqnarray}
 Hence,  $U_{\rm inv}/U_{\rm min}$ is given by
\begin{eqnarray}
 \frac{U_{\rm inv}}{U_{\rm min}} \simeq
 \eta_{\rm c} - 
 0.5 \left\{ \left(\frac{U_{\rm e}}{U_{\rm B}}\right)^{3/7}
          + \left( \frac{U_{\rm e}}{U_{\rm B}}\right)^{-4/7} \right\}.
\label{inv}
\end{eqnarray} 
 Recent observations show that the ratio 
 of $U_{\rm e}$ to $U_{\rm B}$ is 
 $1 \lesssim U_{\rm e} / U_{\rm B} \lesssim 10$  on average.
 Substituting these values in  Eq. (\ref{ratio2}),
 we  find $0.13\lesssim U_{\rm B}/U_{\rm min} \lesssim 0.5$ and
 $0.5\lesssim U_{\rm e}/U_{\rm min} \lesssim 1.3$. 
 Since the obtained range of $\eta_{\rm c}$
 is
 $\sim 4-310$,
  $U_{\rm inv}/U_{\rm min}$ is evaluated as
 $3\lesssim U_{\rm inv}/U_{\rm min} \lesssim 310$
 from Eq. (\ref{inv}).
 Thus, we conclude that
 the internal energy  of invisible particles
 must be larger than  
 those of radiating 
 non-thermal leptons and magnetic fields 
($U_{\rm e}, U_{\rm B} \lesssim U_{\rm inv}$)
 to explain the result obtained in this paper 
 that $E_{\rm c}$ is larger than
 $E_{\rm min}$ by a factor of 4 at least.


\acknowledgments

We are grateful to A. Celotti for useful comments.
We thank M. Machida for useful comments and discussions on \S 5.1.
 This work was partially supported by the Grants-in-Aid for the
 Scientific Research (14740166, 14079202) from Ministry of Education,
 Science and Culture of Japan and by Grants-in-Aid for the 21th century
 COE program ``Holistic Research and Education Center for Physics of
 Self-organizing Systems''.
 This research has made use of SAOimage DS9, developed by Smithsonian
 Astrophysical Observatory.
 
%
%





\begin{table}
\begin{center}
\caption{The quantities measured from observations.\label{tab1}}
\begin{tabular}{cccccccrrrrrrr}
\tableline\tableline

 Source & $r_{\rm h}$ (kpc)
       & $A_{\rm h}$ (kpc$^2$)
       & $\rho_{\rm a}$  (g cm$^{-3}$)
       & $P_{\rm a}$  (dyne cm$^{-2}$)
       & $\alpha$
       & Ref. \\

 (1) & (2)
       & (3)
       & (4)
       & (5)
       & (6)
       & (7)\\
\tableline

 Cygnus A E & 60 & 150  &
 8.3$\times$10$^{-27}$   & 8.0$\times$10$^{-11}$
   & 1.5 & 1,2 \\

 Cygnus A W  & 70 & 150  &  6.6$\times$10$^{-27}$ & 6.4$\times$10$^{-11}$
   & 1.5 & 1,2\\

3C 223 N & 340 & 4300 & 5.5$\times$10$^{-28}$ & 
  1.2$\times$10$^{-12}$  & 1.6 & 3 \\

3C 223 S & 340 & 1800 & 5.5$\times$10$^{-28}$ &  
  1.2$\times$10$^{-12}$  & 1.6 & 3  \\

3C 284 E & 260 & 4600 & 4.0$\times$10$^{-28}$ & 
 6.4$\times$10$^{-13}$  & 1.0 & 3  \\

3C 284 W & 380 & 6200 & 2.3$\times$10$^{-28}$ & 
  3.7$\times$10$^{-13}$   & 1.4 & 3 \\

3C 219 W & 210 & 5000 & 1.0$\times$10$^{-27}$ & 
 1.6$\times$10$^{-12}$  & 2.0 & 4 \\

\tableline
\end{tabular}

\tablecomments{Column (1) shows the names of  radio sources, and the
 following alphabet distinguishes the pair of  jets (see Fig. \ref{radio}
 ). 
 Columns (2) and (3) display, respectively, the cocoon lengths and the cross
 sectional areas of cocoon head  measured from Fig. \ref{radio}
 .
 Columns (4) and (5) give the estimated 
 ICM densities and pressures at $r = r_{\rm h}$.
 In Column(6), the estimated power-law indexes of the ICM density are
 presented.
 References for the density profiles and pressures are listed in column (7)}

\tablerefs{
(1) \citet{RF96}; (2) \citet{SW02}; 
 (3) \citet{CB04};
 (4)\citet{HW99}.}

\end{center}
\end{table}

\begin{table}[ht]
\begin{center}
\caption{The observed radio information.\label{tab2}}
\begin{tabular}{cccccccccrrr}
\tableline\tableline

 Source & z
       & d$_{\rm L}$
       & $F_{\nu}$
       & $\alpha_{\rm R}$ 
       & $L_{\nu}$  \\

  & 
       & (Mpc)
       & (Jy)
       &
       & (ergs s$^{-1}$Hz$^{-1}$) \\

 (1) & (2)
       & (3)
       & (4)
       & (5)
       & (6) \\

\tableline

 Cygnus A   & 0.0565 & 249  & 9660 & 0.74
   & 6.2$\times$10$^{35}$ \\ 

3C 223  & 0.1368 & 635 & 16.0 & 
  0.74  &  7.7$\times$10$^{33}$ \\ 

3C 284  & 0.2394 & 1182 & 12.3 & 
  0.95   &  1.0$\times$10$^{34}$ \\ 

3C 219  & 0.1744 & 829 & 44.9 & 
 0.81  &  3.7$\times$10$^{34}$ \\ 

\tableline
\end{tabular}

\tablecomments{Column (1) shows the names of radio sources. 
 Columns (2) and (3) display, respectively, the redshift and the
 luminosity distance calculated for the cosmology with  $H_0=71$km s$^{-1}$,
 $\Omega_{\rm M}=0.3$, and  $\Omega_{\rm \Lambda}=0.7$.
 Columns (4) and (5) give the values of flux densities and spectral
 indexes at 178MHz, which are taken from the table (Table 1) of \citet{HAP98}.
 In Column (6)  the calculated luminosity densities at 178MHz is presented
 }

\end{center}
\end{table}

\begin{deluxetable}{lcccccccccccrrrrrrr}
\rotate
\tablecaption{The obtained properties of the jet and the cocoon together
 with minimum energy of the radio lobe.}
\tablewidth{0pt}

\startdata
\tableline\tableline

 Source & $L_{\rm j}$ 
       & $t_{\rm age}$ 
       & $M_{\rm BH}$
       & $2 L_{\rm j}/L_{\rm Edd}$   
       & $E_{\rm c}$ 
       & $E_{\rm min}$ 
       & $\eta_{\rm c}$
 \\

  & ($10^{46}$ ergs s$^{-1}$)
       & (Myr)
       & ($M_{\odot}$)
       &
       & ($10^{60}$ ergs)
       & ($10^{60}$ ergs)
       &    \\
       
 (1) & (2)
       & (3)
       & (4)
       & (5) 
       & (6)
       & (7)
       & (8)  \\

\tableline

 Cygnus A E & 0.4 - 2.6 & 19 - 47  & 
      2.5$\times$10$^9$(2) &      
      0.025 - 0.16 &
       6.2 - 16  & 
       1.4 &  
       4.4 - 11   
      \\

 Cygnus A W & 0.35 - 1.1 & 30 - 53  &
      2.5$\times$10$^9$(2) &
      0.021 - 0.068 &
       6.1 - 11  & 
      1.4 &  
       4.3 - 7.8   
      \\

 3C 223 N & 0.15 - 2.9 & 140 - 610  &
      1.4$\times$10$^8$(1) &
      0.16 - 3.2 & 
       30 - 130  & 
      0.88  &
       34 - 150   
      \\

 3C 223 S & 0.071 - 0.2 & 330 - 560  &
      1.4$\times$10$^8$(1) &
      0.078 - 0.22 & 
       12 - 22  & 
      0.88 &
       14 - 25   
      \\


3C 284 E  & 0.3 - 18 & 32 - 260  & 
      8.2$\times$10$^8$(3,4) &
      0.053 - 3.4 & 
       26 - 210  & 
       1.8    &
       14 - 120   
      \\

 3C 284 W &  0.1 - 3.6  & 100 - 630  &
      8.2$\times$10$^8$(3,4) &
      0.018 - 0.67 & 
       21 - 130  & 
      3.0   &
       7 - 43   
      \\

3C 219 W  & 2.6 - 43 & 37  - 150  &
     6.3$\times$10$^8$(3) &
      0.65 - 10 &
       130 - 500  & 
      1.6    &
       79 - 310  & 
       \\

\enddata

\tablecomments{Column (1) shows the names of  radio sources, and the
 following alphabet distinguishes the
 pair of  jets (see Fig. \ref{radio}).
 Columns (2) and (3) display, respectively, the total kinetic powers and 
 ages of the radio jets. In column (4) and (5),  the black hole mass
 and the kinetic powers
 normalized by the corresponding Eddington luminosity are displayed,
 respectively.
 References for the
 central SMBH mass are given in parentheses.
 In Columns (6) and (7), 
 the total energies
 deposited in the cocoon  and
 the minimum energies required for the synchrotron emission 
 are displayed,
 respectively.
 Column (8) gives the ratios between $E_{\rm c}$ and $E_{\rm min}$.}

\tablerefs{
 (1) \citet{WU02}; 
 (2) \citet{TMA03}; 
 (3) \citet{MCF04};
 (4) \citet{SRH05}.}
\label{tab3}
\end{deluxetable}

\end{document}